\documentclass[preprint,12pt]{elsarticle}
\usepackage{array}
\usepackage{mathtools}
\usepackage{amsmath}
\usepackage{algorithm}
\usepackage{algpseudocode}

\usepackage{amssymb}

\usepackage{color}
\usepackage{graphicx}[dvipdfmx]

\newcounter{bla}

\journal{Computer Physics Communications}

\begin{document}

\begin{frontmatter}



\title{102 PFLOPS Lattice QCD quark solver on Fugaku}


\author[a]{Ken-Ichi~Ishikawa}
\author[b]{Issaku~Kanamori}
\author[c]{Hideo~Matsufuru}
\author[d]{Ikuo~Miyoshi}
\author[d]{Yuta~Mukai}
\author[b]{Yoshifumi~Nakamura\corref{author}}
\author[b]{Keigo~Nitadori}
\author[b]{Miwako~Tsuji}

\cortext[author] {Corresponding author.\\\textit{E-mail address:} nakamura@riken.jp}

\address[a]{Core of Research for the Energetic Universe, Graduate School of Advanced Science and Engineering, Hiroshima University, Higashi-Hiroshima 739-8526, Japan}

\address[b]{RIKEN Center for Computational Science, Kobe, Hyogo 650-0047, Japan}

\address[c]{Computing Research Center, High Energy Accelerator Research Organization (KEK) Oho 1-1, Tsukuba 305-0801, Japan}
\address[d]{Fujitsu Limited, Kawasaki, Kanagawa, 211-8588, Japan}

\begin{abstract}
\vskip-9.5cm
\noindent\hfill HUPD-2108
\vskip9.5cm 
We present results on the world's first over 100 PFLOPS single precision lattice QCD quark solver on the japanese new supercomputer Fugaku.
We achieve a factor 38 time speedup from the supercomputer K on the same problem size, $192^4$,
with 102 PFLOPS, 10\% floating-point operation efficiency against single precision floating-point operation peak.
The evaluation region is the single precision BiCGStab for a Clover--Wilson Dirac matrix with 
Schwarz Alternating Procedure domain decomposition preconditioning using Jacobi iteration for the local domain matrix inversion.
\end{abstract}


\end{frontmatter}

\section{Introduction}\label{sec:intro}

Lattice quantum chromodynamics (LQCD) is an application to solve problems in elementary particle physics.
In LQCD, we compute the quantum chromodynamics (QCD) theory of quarks and gluons on the 
four-dimensional (4D) regular lattice. The solver of the Dirac equations, which is called ``quark solver'' and
solved by iterative methods, consumes a large portion of CPU time in this application.

The coefficient matrix of the quark solver using Wilson type fermions is a large sparse matrix.
The 4D space and time is discretized to a 4D square lattice.
Quark field of 12 complex numbers is put at the site and gauge field of 9 complex numbers is put at the link on the lattice. 
Computaional speed is usually limited by the main memory bandwidth and the network bandwidth, although
the required memory size is relatively small compared to other applications.
Therefore achieving nice strong scaling is very challenging in LQCD.

In 2014, the Japanese gorvernment launched a project called ``FLAGSHIP 2020 project'', aimed to develop a new national flagship supercomputer 
covering a wide range of applications that would run on the system to solve societal and scientific issues.
The key concept of the project is ``co-design'' of the system and applications, where the application is optimized for the system, while the system is designed to satisfy the requirements of the application~\cite{Richard2013CoDesign}. 
Using the applications developed for such societal and scientific issues including LQCD, the performance of the system had been estimated to explore the design space such as SIMD-length, cache sizes and so on~\cite{Sato2020CoDesign}. 
LQCD contributes to the co-design as an application requiring high memory bandwidth and network bandwidth,
the communication mechanism of direct memory access, low cache latency, low network latency, no OS jitter,
and enough registers to maximize performance of floating-point arithmetic unit by out-of-order (OoO) execution.
In this context, we describe our advanced optimization of the quark solver for Fugaku and the benchmark results 
of a practical physical problem size with the entire system of Fugaku.

This paper is organized as follows.
In section~\ref{sec:fugaku} we explain the specification of the supercomputer Fugaku (hereinafter referred to as Fugaku).
In section~\ref{sec:solver} 
we describe the structure and algorithms of the quark solver, and the regions to be optimized and measured for the benchmarking on Fugaku.
The details of optimization employed in the quark solver on Fugaku are explained in section~\ref{sec:tuning}.
The results of the benchmark tests on Fugaku are presented in section~\ref{sec:result}.
Our conclusions are summarized in section~\ref{sec:summary}.

\section{Fugaku}\label{sec:fugaku}

Fugaku is a new japanese supercomputer developed by RIKEN and \mbox{Fujitsu}, 
which is the successor of the supercomputer K (hereinafter referred to as K).
In the recent TOP500, HPCG, HPL-AI, and Graph500 benchmark rankings, 
it has been ranked No. 1 in the world for three consecutive terms (June 2020, November 2020, and June 2021)~\cite{top500}.
It has a total of 432 racks with a total of 158976 nodes (384 nodes $\times$ 396 racks $+$ 192 nodes $\times$ 36 racks).
One node has one A64FX processor~\cite{A64FX} and 32 GiB main memory (four 8 GiB High Bandwidth Memory 2).
The peak memory bandwidth per node is 1024 GB/s.
The node has two external interfaces, one is computational network called 
Tofu interconnect D (TofuD)~\cite{TofuD} and another is PCIe Gen3 16 lanes.

The A64FX processor has designed based on Armv8.2-A instruction sets with the Scalable Vector Extension (SVE).
A64FX has 48 cores for compute and two or four assistant cores for OS services. 
It consists of four Core Memory Groups (CMG) connected by ring bus.
Each CMG has 12 compute cores which share L2 cache.
Two operating frequency modes of normal 2.0 GHz and boost 2.2 GHz are available on Fugaku. 
While the specification of the Armv8.2-A with SVE allows hardware developers to select a vector length from 128 to 2,048 bits, the 512-bit-width SIMD arithmetic units had been chosen on Fugaku.
The processor supports OoO execution of instructions. 

\begin{figure}[htbp]
  \begin{center}
  \includegraphics[width=\textwidth, bb=0 0 810 843]{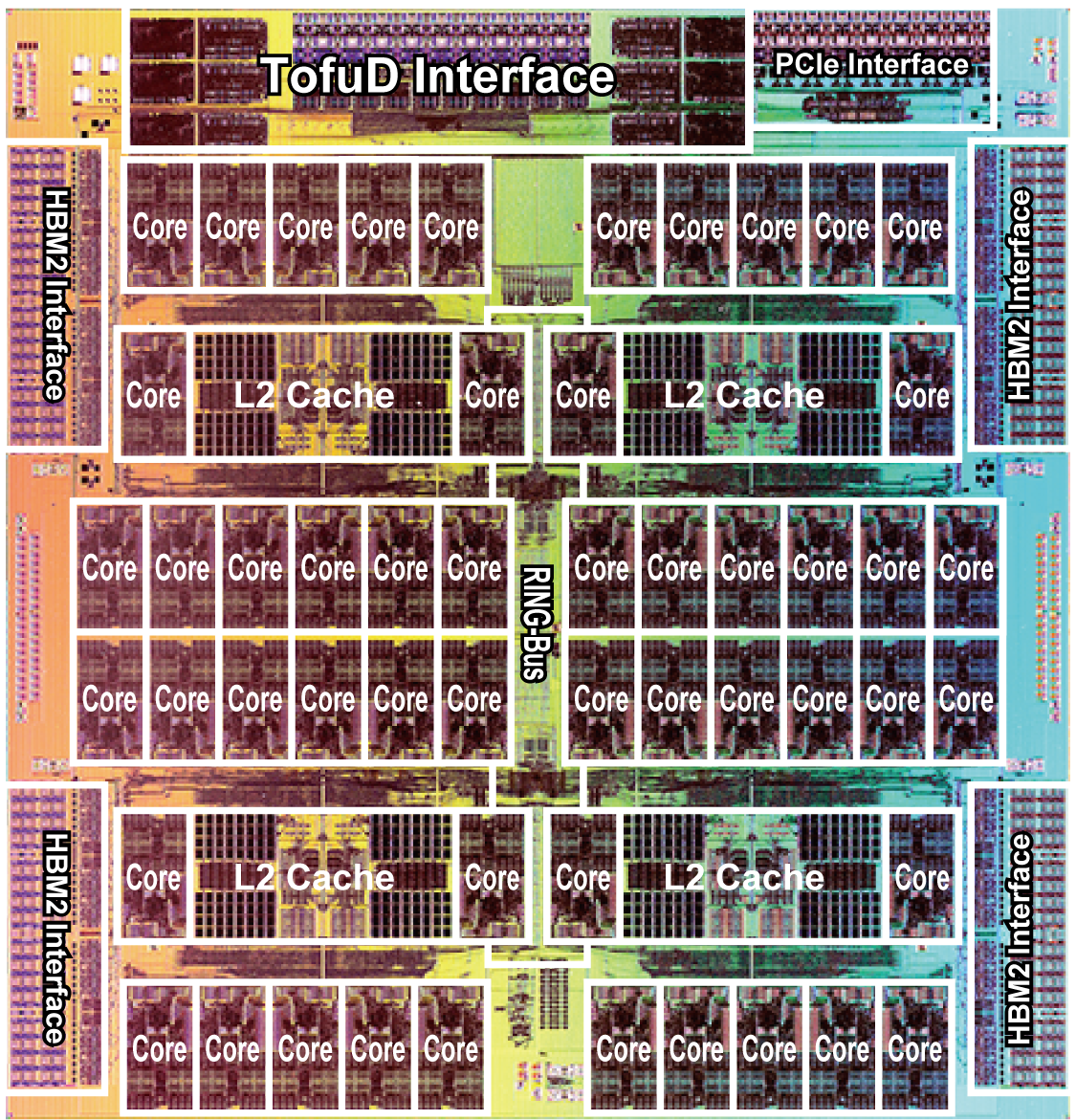}
  \end{center}
  \caption{Die picture of A64FX.}
  \label{fig:a64fx}
\end{figure}

Other specifications are summarized in Table~\ref{tab:a64fx}.
The CPU die picture of A64FX is shown in Fig.~\ref{fig:a64fx}.

\begin{table}[htbp]
   \centering
   \caption{A64FX CPU Specifications. TF denotes TFLOPS. Cache performance is at normal mode of 2.0 GHz.}
   \label{tab:a64fx}
   \begin{tabular}{l | l | l | l} \hline
      & \multicolumn{3}{l}{description}  \\\hline
     Architecture & \multicolumn{3}{l}{Armv8.2-A SVE (512 bit SIMD)}  \\\hline
     Core & \multicolumn{3}{l}{48 ($+$ 2 or 4 assistant cores)} \\\hline
     Perfromance & double prec. & single prec. & half prec. \\ 
     Normal mode & 3.072 TF  &6.144 TF &12.288 TF \\
     Boost  mode & 3.3792 TF &6.7584 TF&13.5168 TF \\\hline
     Cache & \multicolumn{3}{l}{L1D/core: 64 KiB, 4way, 256 GB/s (load), 128 GB/s (store)} \\
              & \multicolumn{3}{l}{L2/CMG: 8 MiB, 16way} \\
              & \multicolumn{3}{l}{L2/node: 4 TB/s (load), 2 TB/s (store)} \\
              & \multicolumn{3}{l}{L2/core: 128 GB/s (load), 64 GB/s (store)} \\\hline
     Memory & \multicolumn{3}{l}{32 GiB, 1024 GB/s} \\\hline
     Interconect & \multicolumn{3}{l}{TofuD (28 Gbps $\times$ 2 lane $\times$ 10 port)} \\\hline
     PCIe & \multicolumn{3}{l}{Gen3 16 lanes} \\\hline
     Technology & \multicolumn{3}{l}{7nm FinFET} \\\hline
\end{tabular}
\end{table}

We have developed ``Lattice quantum chromodynamics simulation library for Fugaku
with wide SIMD'' (QWS~\cite{QWS}) to get high performane for computing quark solver on Fugaku.
It is an open source software on github.
The QWS library contains not only the optimized linear solver for quark solver but also may functions
such as the sparse-matrix-vector multiplication routines or linear-algebra routines for quark fields, 
which can be used as  building blocks of algorithms developed by the library users.
As a practical example, we introduce the use in the multi-grid
solver algorithm implemented in the Bridge++ code set in which
the domain-decomposed Clover-Wilson operator in QWS is called
alternatively to the original code and achieves considerable
acceleration~\cite{Kanamori2021}.
While the convention of the gamma-matrix and data layout are
different in QWS and Bridge++, they are properly converted
before and after the QWS functions are called.

We also expect that QWS provides a working example of the
optimization techniques for A64FX architecture.
While the quantitative evaluation of the components of QWS is
specific to the fermion operator and execution setup,
the improved implementation is generic and applicable to
the other kinds of fermion operators as well as to many
ingredients of LQCD simulations.
Indeed in the multi-grid solver in~\cite{Kanamori2021} the other
ingredients, such as the fermion operator on the coarse lattice
and the inter-process communication, are accelerated by partly
referring to the QWS implementation.

\section{Quark solver on Fugaku}\label{sec:solver}
%
%
We consider to solve a linear equation derived from the equation of motion of quarks
by the lattice descretization of the 4D space-time.
We employ the so-called Clover--Wilson Dirac lattice quark action for the equation of motion of quarks on the lattice.
In LQCD, we have to solve the following linear equation
\begin{align}
  A x = b,
\label{eq:basiceq}
\end{align}
by a huge amount of times using the quark solver.
The coefficient matrix $A$ derived from  the Clover--Wilson lattice quark action is called
the Clover--Wilson Dirac operator in the literature.

The explicit structure of the matrix-vector multiplication $A x$ is as follows.
\begin{align}
&  \sum_{\beta=1}^{4}\sum_{b=1}^3\sum_{m}A_{(a,\alpha),(b,\beta)}(n,m) x_{(b,\beta)}(m)
    \notag\\
&  \quad
    = x_{(a,\alpha)}(n)
    - \kappa
  \sum_{\beta=1}^{4}\sum_{b=1}^3
  \sum_{\mu=1}^{4}\left[
               \left(C_{\mathrm{inv}}T^{+}_{\mu}\right)_{(a,\alpha),(b,\beta)}(n) x_{(b,\beta)}(n+\hat{\mu})
               \right.
  \notag\\
&  \qquad  \qquad  \qquad  \qquad  \qquad  \qquad  \qquad
  \left.
     +\left(C_{\mathrm{inv}}T^{-}_{\mu}\right)_{(a,\alpha),(b,\beta)}(n) x_{(b,\beta)}(n-\hat{\mu})\right],
\end{align}
where $n$ and $m$ denote the 4D lattice site indices, $a$ and $b$ are color
indices running from 1 to 3, and $\alpha$ and $\beta$ are spin indices from 1 to 4.
$x_{(a,\alpha)}(n)$ is a quark field element with the color index $a$ and spin index $\alpha$
located at the lattice site $n$ and takes a complex number.
$\hat{\mu}$ denotes a unit vector pointing $\mu$-th direction in the 4D lattice.
The small coefficient matrices ${C_{\mathrm{inv}}T^{\pm}_{\mu}}$
represent the local dynamics among quarks and gluons by acting on the 12 complex components of the quark field,
and vary during LQCD simulations.
The details of ${C_{\mathrm{inv}}T^{\pm}_{\mu}}$ are given in~\ref{sec:dirac}.
$\kappa$ parametrizes the quark mass, the smaller the heavier and the larger the lighter.
Suppressing indices and its summation for simplicity, we use the following notation
\begin{align}
  A x = \left[I - \kappa C_{\mathrm{inv}}H \right] x,
  \label{eq:hoppingmat}
\end{align}
where $C_{\mathrm{inv}}H$ is called the hopping matrix and defined as
the collection of the 4D three point stencil acting on
the 12 components of the quark vector.

To solve Eq.~\eqref{eq:basiceq}, we employ the BiCGStab algorithm.
Directly solving Eq.~\eqref{eq:basiceq} in double precision is, however, not efficient due to the following mathematical and engineering reasons:
incorporating preconditioning is quite recommended for iterative algorithms for numerical stability and efficiency,
and current supercomputer systems are more powerful with lower precision arithmetic. 
We, therefore, employ the Schwartz Alternating Procedure (SAP) preconditioning with domain-decomposition~\cite{SAP}
combined with the mixed-precision iteration technique~\cite{MixedPrecision,MixedPrecision2,MixedPrecision3}.
In the following, we explain the structure of the SAP preconditioning and
the mixed-precision solver technique implemented in QWS for Fugaku.

\begin{figure}[htbp]
  \begin{center}
  \includegraphics[width=\textwidth, bb= 0 0 593 341]{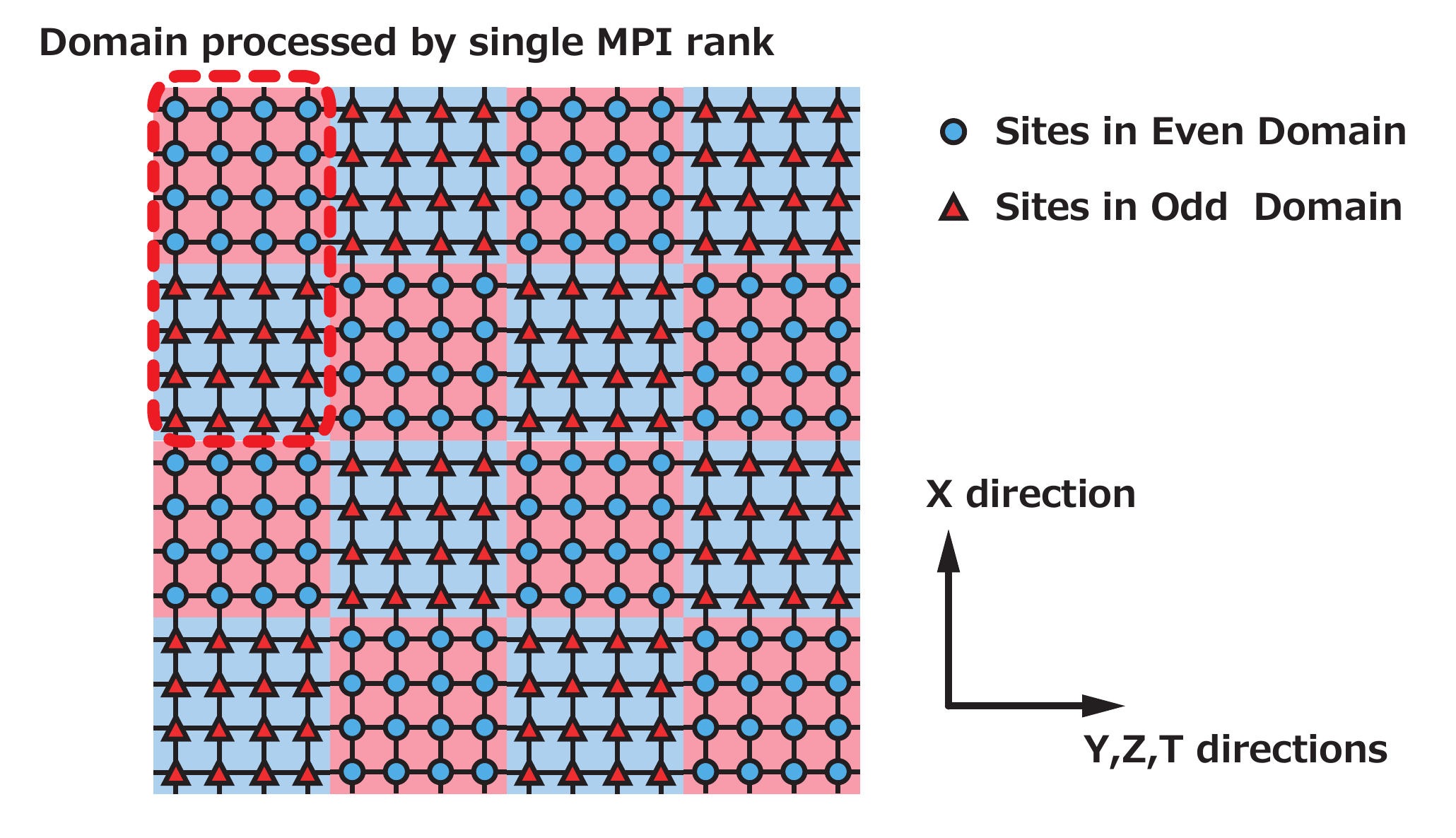}
  \end{center}
  \caption{Sketch of domain decomposition.}
  \label{fig:domaindecomposition}
\end{figure}

The SAP preconditioning is an approximation of the block Gauss-Seidel iteration. In our implementation on Fugaku,
the block corresponds to the non-overlapping domain-decomposition of the lattice volume.
We divide the whole lattice volume into non-overlapping domains as shown in Fig.~\ref{fig:domaindecomposition}.
Attached two colors (even or oddness of the domain indices) to each domain and using block notation,
the coefficient matrix $A$ and vector $x$ can be expressed as
\begin{align}
A =   \begin{bmatrix*}[c]
 A_{EE} & A_{EO} \\
 A_{OE} & A_{OO} \\
\end{bmatrix*},  \qquad
  x =
  \begin{bmatrix*}[c]
      x_{E} \\ x_{O}
  \end{bmatrix*},
\end{align}
where subscripts $E$ and $O$ denote the site indices belong to the even and odd domains, respectively.
$A_{EE}$ only acts on $x_{E}$ located on even-colored domain sites, while
$A_{EO}$ maps a vector $x_{O}$ located on odd-colored domain to a vector on even-colored domain.
From the three-point stencil structure of $A$, $A_{EO}$ ($A_{OE}$) are the stencil operation from
the surface sites of the nearest neighbor odd (even) domain to the ones of the even (odd) domain, respectively.

The SAP preconditioner $M_{\mathrm{SAP}}$ is defined as
\begin{align}
  M_{\mathrm{SAP}} &= K \sum_{j=0}^{N_{\mathrm{SAP}}-1} (1-A K)^{j},\notag\\
  K&= \begin{bmatrix*}[c]
          B_{EE} & O \\
         -B_{OO} A_{OE} B_{EE} & B_{OO}
     \end{bmatrix*},
   \notag\\
  B_{EE} & \simeq A_{EE}^{-1},\qquad
  B_{OO}  \simeq A_{OO}^{-1},
\label{eq:Msap}
\end{align}
where $B_{EE}$ ($B_{OO}$) represents an approximation for $A_{EE}^{-1}$ ($A_{OO}^{-1}$).
$N_{\mathrm{SAP}}$ is an integer parameter of the SAP iteration.
When $B_{EE}$ ($B_{OO}$) is exact to $A_{EE}^{-1}$ ($A_{OO}^{-1}$), the matrix $AK$ becomes upper triangular as
\begin{align}
  AK
  &=  \begin{bmatrix*}[c]
          A_{EE}  & A_{EO} \\
          A_{OE}  & A_{OO}
      \end{bmatrix*}
      \begin{bmatrix*}
          A_{EE}^{-1}  & O \\
         -A_{OO}^{-1}A_{OE}A_{EE}^{-1}  & A_{OO}^{-1}
      \end{bmatrix*}
                                          \notag\\
  &=  \begin{bmatrix*}
        I -A_{EO}A_{OO}^{-1}A_{OE} A_{EE}^{-1} & A_{EO}A_{OO}^{-1} \\
          O  & I
      \end{bmatrix*}.
\end{align}
Therefore it is expected the matrix $AK$ has a smaller condition number than that of $A$
even with approximation $B_{EE}\simeq A_{EE}^{-1}$ ($B_{EE}\simeq A_{EE}^{-1}$).
The SAP preconditioner $M_{\mathrm{SAP}}$ \eqref{eq:Msap} is an approximation for $A^{-1}$
followed by a truncated Neumann series for $(AK)^{-1}$ with the preconditioned matrix $AK$.

The approximate inverse matrix $B_{EE}$  are evaluated with the truncated Neumann
series or equivalently Jacobi iteration as
\begin{align}
  B_{EE}
  &= \sum_{j=0}^{N_{\mathrm{JAC}}} (I - A_{EE})^{j}.
\end{align}
The same form holds for $B_{OO}$ by replacing $EE$ with $OO$.

To achieve an efficient parallel performance, we divide the whole lattice volume consistent with
the SAP preconditioning as shown in Fig.~\ref{fig:domaindecomposition}.
We include two neighboring domains on an MPI rank, the one is even-colored and the other is odd-colored,
so that all the MPI ranks participate in computing either of $B_{EE}$ or $B_{OO}$.
On the other hand, in computing $A_{EO}$ or $A_{OE}$, it always requires MPI communication
to do nearest neighbor stencil computation between surface sites in opposite colored domains.
To minimize the communication time overhead, we overlap the point-to-point communication to
the computation related to $A_{EE}$ or $A_{BB}$.
We will explain the details of the communication-computation overlapping using Fugaku's Tofu hardware
and software in section~\ref{sec:tuning}.

The algorithms for matrix vector multiplication $x = M_{\mathrm{SAP}}b$ and $x_E = B_{EE} b_E$ are
shown in Algorithms~\ref{alg:qws_prec_s} and \ref{alg:qws_jinv_ddd_in_}, respectively.
The communication-computation overlapping is possible for $M_{\mathrm{SAP}}$ as shown in
lines \ref{alg:line:send:xE} and \ref{alg:line:send:xO} of Algorithm~\ref{alg:qws_prec_s}.
However, the last communication (line~\ref{alg:line:send:xElast}) cannot be overlapped.
To further overlap the last communication over the computation, we also provide $A M_{\mathrm{SAP}}$ as a combined
matrix-vector multiplication routine for the iterative solver. The algorithm for $x = AM_{\mathrm{SAP}} b$ is shown in Algorithm~\ref{alg:qws_prec_ddd_s},
where all communications are overlapped with computations.

Provided the above matrix-vector multiplication routines (Algs.~\ref{alg:qws_prec_s}, \ref{alg:qws_jinv_ddd_in_}, and -\ref{alg:qws_prec_ddd_s})
as building blocks, instead of solving Eq.~\eqref{eq:basiceq} directly,
we apply the BiCGStab algorithm to solve
\begin{align}
  (A M_{\mathrm{SAP}}) y = b,
\label{eq:SAPprecEQ}
\end{align}
for $y$ and the solution for $x$ is given by
\begin{align}
 x = M_{\mathrm{SAP}}y.  
\end{align}

In this paper, we focus on the performance tuning details of the single-precision BiCGStab solver for solving Eq.~\eqref{eq:SAPprecEQ}.
However, realistic LQCD simulations need the double-precision (DP) solutions of Eq.~\eqref{eq:basiceq}.
To obtain the double-precision solution using single-precision (SP) arithmetic as far as possible,
we employ the so-called mixed-precision strategy~\cite{MixedPrecision}.
Here we briefly explain the idea of the strategy before explaining the details of the single-precision BiCGStab
algorithm implemented in QWS for Fugaku.

The key idea of the mixed-precision is in the Richardson-iteration or equivalently
deficit correction iteration method with the approximate single-precision solutions.
When an approximation $x_{\mathrm{SP}} = A_{\mathrm{SP}}^{-1} b_{\mathrm{SP}}$ with SP is available,
the backward error in DP for Eq.~\eqref{eq:basiceq} is
\begin{align}
  r = b - A P_{DP}(x_{\mathrm{SP}}),
\end{align}
where $r,b$ and $A$ are in DP, and it is omitted for clarity.
The operation $P_{DP}(\dots)$ converts the precision of the argument from SP to DP.
The DP solution $x$ requires $r=0$ in DP, we can enforce this by adding a correction vector $\delta x$
to $P_{DP}(x_{\mathrm{SP}})$ as
\begin{align}
  x = P_{DP}(x_{\mathrm{SP}}) + \delta x.
  \label{eq:corrected}
\end{align}
The vanishing residual for $x$ in Eq.~\eqref{eq:corrected} requires
\begin{align}
  0 & = b - A x \notag\\
    & = b - A(P_{DP}(x_{\mathrm{SP}}) + \delta x)
      \notag\\
    &= b -AP_{DP}(x_{\mathrm{SP}}) - A \delta x\notag\\
  &= r - A \delta x.
\end{align}
The correction vector $\delta x$ must be the solution of $A \delta x=r$ and it is again obtained
approximately with a SP iterative solver as $\delta x_{\mathrm{SP}} = A^{-1}_{\mathrm{SP}}r_{\mathrm{SP}}$.
This process repeats until the residual converges below a tolerance in DP.

Algorithm~\ref{alg:DCI} shows the pseudo-code. $R_{\mathrm{SP}}(\dots)$ reduces the precision of the argument from DP to SP.
The rescalings with err at lines \ref{alg:line:norma1} and \ref{alg:line:norma2} are required
to avoid underflow in the single precision solver at the line \ref{alg:line:spsolve}.
This DP iteration loop (lines \ref{alg:loopbegin}--\ref{alg:loopend}) is called outer-loop
while the iterative loops in the SP solver (line 6) are inner-loop.
We can employ more sophisticated algorithms for the outer-loop where the SP solver (inner-loop) is regarded 
as a flexible preconditioner to the target linear equation.
The ``flexible'' means that the preconditioner can be flexibly varied iteration
by iteration during the outer-loop iterations.
Flexible GMRES, Flexible BiCG and BiCGStab are examples of the flexible iterative
solver algorithms for non-Hermitian matrices~\cite{flexiblesolver}.
As seen from Algorithm~\ref{alg:DCI},
it is expected that the residual norm $|r|$ reduces by a factor of $O(10^{-6})$ by every iteration
provided that the SP solution has an accuracy of $|p_{SP}-A_{SP}q_{SP}|< O(10^{-6})$,
and the DP solution will be obtained after a few iterations of the outer-loop.

Algorithm~\ref{alg:qws_bicgstab_precdd_s} shows the single-precision BiCGStab solver to solve Eq.~\eqref{eq:SAPprecEQ} in QWS on Fugaku.
In our implementation, we employ a slightly different algorithm from those in the literature.
The iteration has two loop-exit points (lines \ref{alg:exit1} and \ref{alg:exit2}), saving
the cost of one matrix-vector multiplication at the convergence.
There are several global synchronization points for the norm and dot-product computations.
To reduce the number of synchronization points, we move the computation of
$\rho=\langle\tilde{r}|r\rangle$ from just before line \ref{alg:rho1old} to line \ref{alg:rho1}.
As we will explain in the next section, Fugaku's TofuD has a capability to 
reduce three floating-point elements simultaneously in a single reduction operation.
In our implementation, therefore, each latency of the global reduction communication is identical
to that of the single element reduction communication.

\makeatletter
\renewcommand{\ALG@beginalgorithmic}{\footnotesize}
\makeatother
\algrenewcommand\algorithmicrequire{\textbf{Input:}}
\algrenewcommand\algorithmicensure{\textbf{Output:}}

\begin{algorithm}[H]
\caption{QWS SAP preconditioner: $x = M_{SAP}b$, $M_{SAP}=$ \texttt{prec\_s\_} $\simeq A^{-1}$. 
Input and output vectors: $b,x$, and working vectors : $s,q$}
\label{alg:qws_prec_s}
\begin{algorithmic}[1]
    \Require{$b$ : source vector,  $N_{\mathrm{SAP}}$, $N_{\mathrm{JAC}}$ : iteration count}
    \Ensure{$x$ : multiplied by the SAP preconditioner $x = M_{SAP} b$}
  \State{$s = b$}
  \For{$\mathrm{i}=0; \mathrm{i} < N_{\mathrm{SAP}}; \mathrm{i}\!+\!+$}
  \State{$ x_E = B_{EE} s_E$}   \algorithmiccomment{[Algorithm~\ref{alg:qws_jinv_ddd_in_}]}
  \State{[Start sending surface data of $x_E$ for $A_{OE}x_E$.] \label{alg:line:send:xE}}
  \State{$ q_E = A_{EE} x_E$}
  \State{$ s_E = s_E + b_E - q_E$}
  \State{[Wait for receiving surface data of $x_E$ from line~\ref{alg:line:send:xE}.]}
  \State{$ s_O = s_O - A_{OE} x_E$}
  \State{$ x_O = B_{OO} s_O$}  \algorithmiccomment{[Algorithm~\ref{alg:qws_jinv_ddd_in_}]}
  \State{[Start sending surface data of $x_O$ for $A_{EO}x_O$.] \label{alg:line:send:xO}}
  \State{$ q_O = A_{OO} x_O$}
  \State{$ s_O = s_O + b_O - q_O$}
  \State{[Wait for receiving surface data of $x_O$ from line~\ref{alg:line:send:xO}.]}
  \State{$ s_E = s_E - A_{EO} x_O$}
  \EndFor
  \State{$x_E = B_{EE} s_E$}  \algorithmiccomment{[Algorithm~\ref{alg:qws_jinv_ddd_in_}]}
  \State{[Start sending surface data of $x_E$ for $A_{OE}x_E$.] \label{alg:line:send:xElast}}
  \State{[Wait for receiving surface data of $x_E$ from line~\ref{alg:line:send:xElast}.]}
  \State{$s_O = s_O - A_{OE} x_E$}
  \State{$x_O = B_{OO} s_O$}  \algorithmiccomment{[Algorithm~\ref{alg:qws_jinv_ddd_in_}]}
\end{algorithmic}
\end{algorithm}

\begin{algorithm}[H]
 \caption{QWS approximate inversion of $A_{EE}$ :
  $x_E=B_{EE}b_E$, $B_{EE}=$ \texttt{jinv\_ddd\_in\_s\_} $\simeq (A_{EE})^{-1}$.
Input and output vectors : $b_E, x_E$, and working vectors $q_E$.}
\label{alg:qws_jinv_ddd_in_}
\begin{algorithmic}[1]
    \Require{$b_E$ : source vector, and $N_{\mathrm{JAC}}$ : iteration count}
    \Ensure{$x_E$ : approximate solution vector, $x_E = B_{EE} b_E$}
  \State{$q_E = A_{EE} b_E$}
  \State{$x_E = 2 b_E - q_E$}
  \For{$\mathit{j}=1; \mathit{j} < N_{\mathrm{JAC}}; \mathit{j}\!+\!+$}
  \State{$ q_E = A_{EE}x_E$}
  \State{$ x_E = x_E + b_E - q_E$}
  \EndFor
\end{algorithmic}
\end{algorithm}

\begin{algorithm}[H]
\caption{QWS multiplying preconditioned matrix : $x=AM_{SAP}b$, $AM_{SAP}=$ \texttt{prec\_ddd\_s\_} $\simeq I$. 
$AM_{SAP}$ is Domain-decomposition preconditioned operator. Input and output vectors : $b,x$, and working vectors : $s,q$.}
\label{alg:qws_prec_ddd_s}
\begin{algorithmic}[1]
   \Require{$b$ : source vector, $N_{\mathrm{SAP}}$, $N_{\mathrm{JAC}}$ : iteration count for Algorithms \ref{alg:qws_prec_s} and \ref{alg:qws_jinv_ddd_in_}.}
   \Ensure{$x$ : multiplied by preconditioner,  $x = AM_{SAP} b$}
  \Statex{[Lines: 1--15 are the same as those of Algorithm~\ref{alg:qws_prec_s}.]}
  \setcounter{ALG@line}{15}
  \State{$ q_E = A_{EE} s_E$}
  \State{[Start sending surface data of $q_E$ for $A_{OE}q_E$.] \label{alg:line:sendDMsaplast}}
  \State{$ x_E = A_{EE} q_E$}
  \State{[Wait for receiving surface data of $q_E$ from line~\ref{alg:line:sendDMsaplast}.]}
  \State{$ s_O = s_O - A_{OE} q_E$}
  \State{$ q_O = A_{OO} s_O$}
  \State{[Start sending surface data of $q_O$ for $A_{EO}q_O$.] \label{alg:line:sendDMsaplast2}}  
  \State{$ x_O = x_O + A_{OO} q_O$}
  \State{[Wait for receiving surface data of $q_O$ from line~\ref{alg:line:sendDMsaplast2}.]}  
  \State{$ x_E = x_E + A_{EO} q_O$}
\end{algorithmic}
\end{algorithm}

\begin{algorithm}[H]
 \caption{Deficit correction iteration to obtain DP solution using SP iterative solver.}
\label{alg:DCI}
\begin{algorithmic}[1]
    \Require{$b$ : DP source vector,  $\mathrm{tol}$ : stopping condition}
    \Ensure{ $x$ : DP solution vector, $x = A^{-1} b$ with $|b-Ax|/|b| < \mathrm{tol}$}
  \State{$ x = 0$}
  \State{$ r = b$}
  \State{$ \mathrm{err} = |r|$}  
  \Loop \label{alg:loopbegin}
  \State{$ p_{\mathrm{SP}} = R_{\mathrm{SP}}(r/\mathrm{err})$ \label{alg:line:norma1}}
  \State{$ q_{\mathrm{SP}} = A_{\mathrm{SP}}^{-1} p_{\mathrm{SP}}$ \label{alg:line:spsolve}}
    \algorithmiccomment{[an iterative solver]}
  \State{$ \delta x = \mathrm{err}\, P_{\mathrm{DP}}(q_{\mathrm{SP}})$ \label{alg:line:norma2}}
  \State{$ x = x + \delta x$}
  \State{$ r = b - A x$ or $r = r - A \delta x$}
  \State{$ \mathrm{err} = |r|$}
  \If{$\mathrm{err}/|b|< \mathrm{tol}$}
    \State{Exit loop}
  \EndIf
  \EndLoop\label{alg:loopend}
\end{algorithmic}
\end{algorithm}

\begin{algorithm}[H]
\caption{QWS $x=(AM_{SAP})^{-1}b$, \texttt{bicgstab\_precdd\_s\_}. Single precision BiCGStab solver for $AM_{SAP}x=b$. Suffixes SP are omitted.
Input and output vectors $b,x$, working vectors $p,q,r,\tilde{r},t$.}
\label{alg:qws_bicgstab_precdd_s}
\begin{algorithmic}[1]
  \Require{$b$ : source vector,  $N_{\mathrm{SPA}}, N_{\mathrm{JAC}}$ : iteration count, $\mathrm{tol}$ : stopping condition}
  \Ensure{$x$ : solution vector,  $x = (DM_{SAP})^{-1} b$ with $|b-DM_{SAP}x|/|b|< \mathrm{tol}$}
  \State $x = 0$
  \State $r = b$
  \State $p = b$
  \State $\tilde{r} = b$   \algorithmiccomment{[1-real reduction = 1 float reduction]}
  \State $\mathrm{bnorm2}=|b|^2$
  \State $\rho_0 = \mathrm{bnorm2}$
  \For{$\mathrm{iter}=0; \mathrm{iter} < \mathrm{maxiter}; \mathrm{iter}\!+\!+$}
  \State $ q = (DM_{SAP}) p $  \algorithmiccomment{[Algorithm~\ref{alg:qws_prec_ddd_s}]}
  \State $ \alpha = \rho_0/ \langle \tilde{r}|q\rangle$ \algorithmiccomment{[1-complex reduction = 2 float reductions]}
  \State $ x = x + \alpha p$
  \State $ r = r - \alpha q$
  \State $ \mathrm{rnorm2} = |r|^2$  \algorithmiccomment{[1-real reduction = 1 float reduction]}
  \If{$\sqrt{\mathrm{rnorm2}/\mathrm{bnorm2}} < \mathrm{tol}$}
  \State exit \label{alg:exit1}
  \EndIf
  \State $ t = (DM_{SAP})r$  \algorithmiccomment{[Algorithm~\ref{alg:qws_prec_ddd_s}]}
  \State $ \omega = \langle t|r\rangle/|t|^2$ \algorithmiccomment{[1-complex + 1-real reductions = 3 float reductions]}
  \State $ x = x + \omega r$
  \State $ r = r - \omega t$
  \State $\mathrm{rnorm2} = |r|^2$, $\rho = \langle \tilde{r}|r\rangle$ \algorithmiccomment{[1-real and 1-complex reductions = 3 float reductions]}\label{alg:rho1}
  \If{$\sqrt{\mathrm{rnorm2}/\mathrm{bnorm2}} < \mathrm{tol}$}
  \State exit \label{alg:exit2}
  \EndIf
  \State $\beta = \alpha\rho/(\rho_0 \omega)$\label{alg:rho1old}
  \State $\rho_0 = \rho$
  \State $p = r + \beta (p - \omega q)$
  \EndFor
\end{algorithmic}
\end{algorithm}

\section{Tuning quark solver on Fugaku}\label{sec:tuning}

\begin{figure}[htbp]
  \begin{center}
  \includegraphics[width=\textwidth, bb= 0 0 1849 1287]{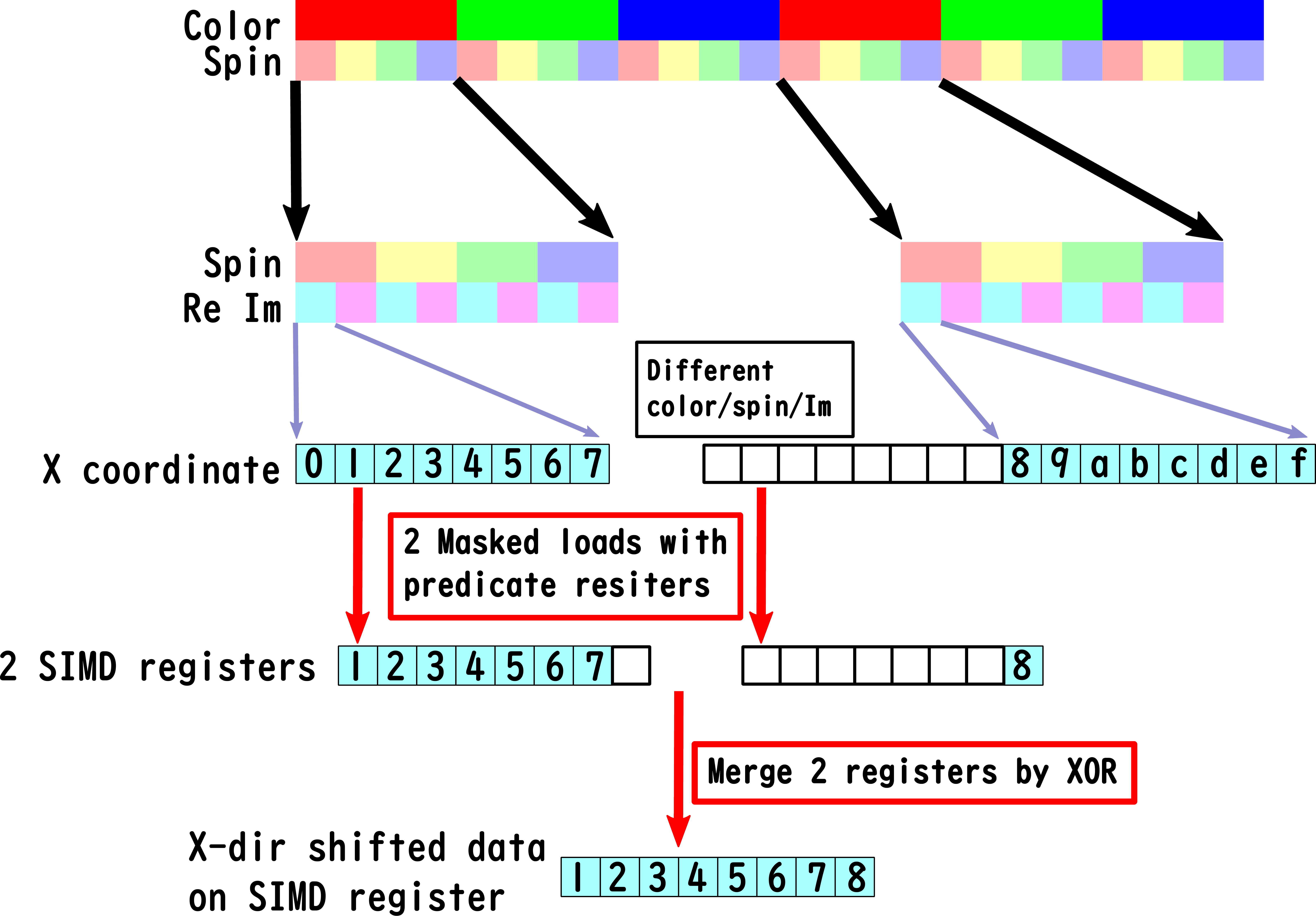}
  \end{center}
  \caption{X--direction shift by using two load operations with 
predicate registers and one XOR operation for data layout in double precision.}
  \label{fig:xdir}
\end{figure}

To get high performance on Fugaku, effective SIMD vectorization with 512 bits wide SIMD is very important.
Since the matrix-vector multiplication of LQCD, which is nine-point stencil type on 4D Euclidean space-time, has a less local complexity using complex arithmetics,
we have to locate single precision data appropriately fit in 512bits wide SIMD vector along with the physics local structure.  We employ the following real number data layout (C-style array or Row major order),
\begin{equation}
\begin{split}
\rm Fugaku(double) &: \rm [nt][nz][ny][nx/8][3][4][2][8]\,,\\
\rm Fugaku(single) &: \rm [nt][nz][ny][nx/16][3][4][2][16]\,,\\
\rm Fugaku(half) &: \rm [nt][nz][ny][nx/32][3][4][2][32]\,,\\
\rm cf.~K    &: \rm [nt][nz][ny][nx][3][4][2]\,,\\
\end{split}
\end{equation}
where ${\rm nt,nz,ny,nx}$ are the local domain lattice size in t,z,y,x directions, respectively. 
{\rm nx} is divided and packed to the SIMD of Fugaku.
The factor 3 sized rank corresponds to the color index, the 4 sized rank to the spin index, and the 2 sized rank to the complex real-imaginary index.
We used the complex number data major layout on K. For Fugaku, we layout  continuous x site index first by blocking to fit with 512 bits wide SIMD vector. 
We optimize the x--direction calculation by using Arm C Language Extensions (ACLE)~\cite{ACLE}.
For the stencil computation in the x--direction, the vector element shift operation is required. Instead of shifting data on vector registers, we utilize vector load with mask operation (named predicate operation) functionality of SVE.
Fig~\ref{fig:xdir} is a schematic picture of the x--direction shift for double precision data layout
by using two load operations with predicate registers and one XOR operation.

We also have applied other general optimizations, e.g., removing temporal arrays,
manually prefetching explicitly 256 B for all arrays by software prefetch,
OMP Parallel region expansion that we put ``omp parallel'' on higher level caller routines because making omp parallel region is costly. 
This is important on many core architectures.

\begin{figure}[!tb]
\centering
\includegraphics[width=10cm, bb= 0 0 595 842]{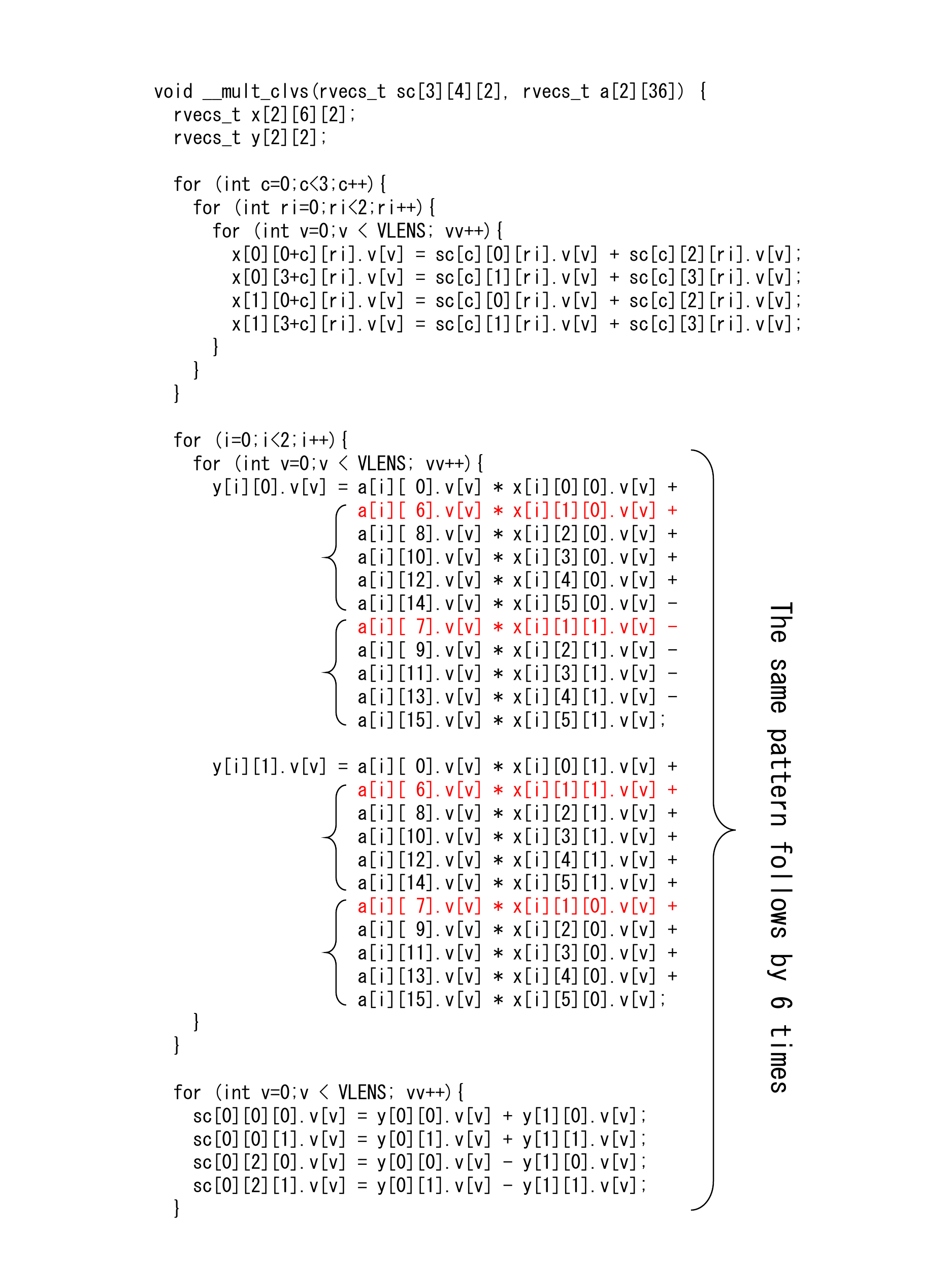}
\caption{\texttt{\_\_mult\_clvs} before optimizations.}
\label{fig:mult_clvs_origcode}
\end{figure}

\begin{figure}[!tb]
\centering
\includegraphics[width=10cm,angle=270, bb= 0 0 540 720]{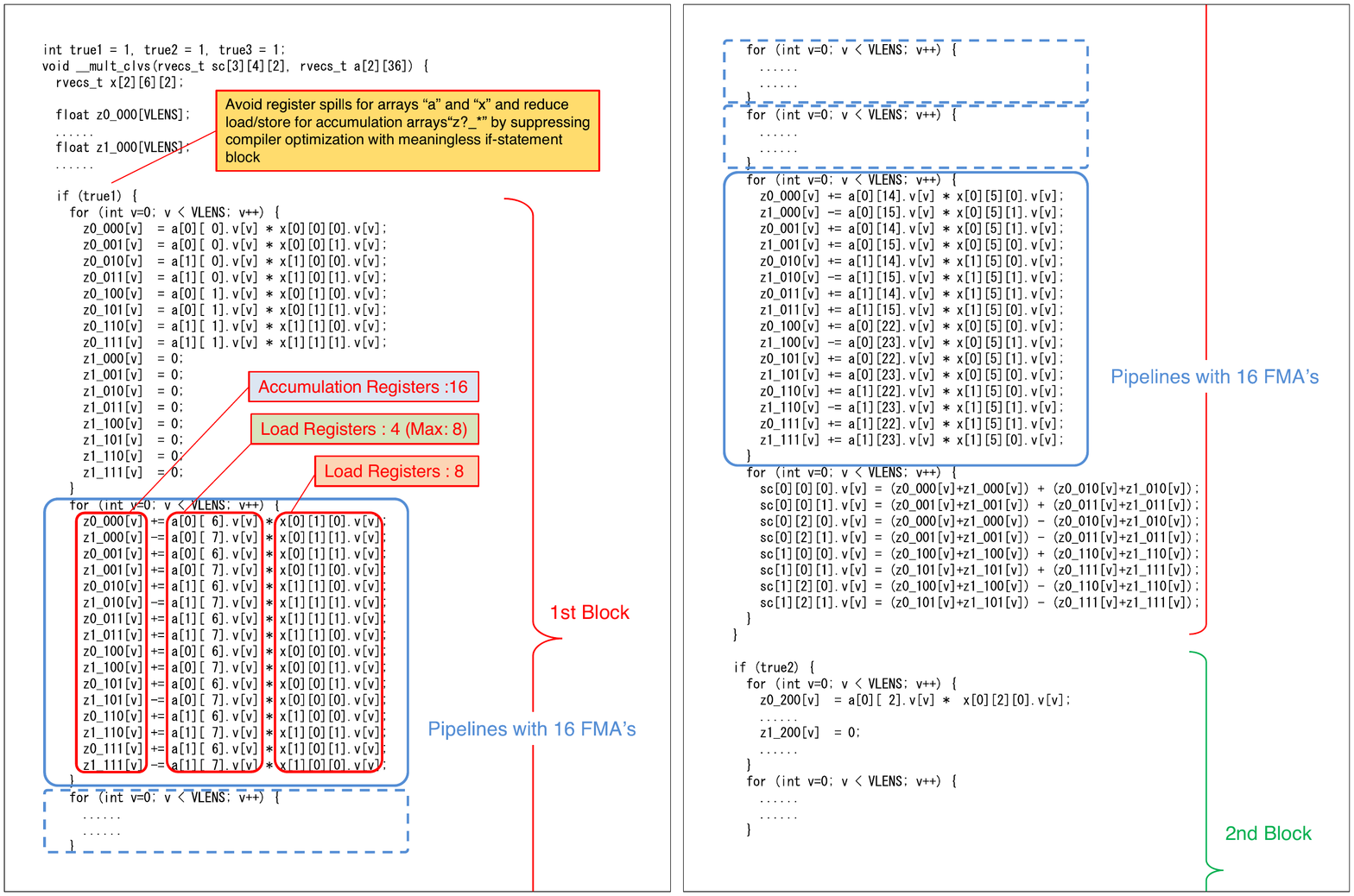}
\vspace{1cm}
\caption{\texttt{\_\_mult\_clvs} after optimizations.}
\label{fig:mult_clvs_modcode}
\end{figure}

A function \verb|__mult_clvs|, which corresponds to local matrix multiplication of $C_\mathrm{inv}$ in Eq.~\eqref{eq:hoppingmat}, 
 is used in \verb|ddd_in_s_|  ($A_{EE}$ or $A_{OO}$), \verb|jinv_ddd_in_s_|  ($B_{EE}$ or $B_{OO}$), and \verb|ddd_out_s_|  ($A_{EO}$ or $A_{OE}$).
We found that a naive implementation of \verb|__mult_clvs| was inefficient due to 
load and store operations caused by register spill/fill. 
It is difficult for the compiler to calculate the optimal instruction schedule for such a large computation.
However, we found a pattern on the code that allows an efficient schedule. Then, we reordered the operations at the source code level.
Fig.~\ref{fig:mult_clvs_origcode} is the outline of the original code. The blocks starting with the red line share four values.
So by arranging the rows of these blocks in a round robin fashion, we can minimize the interval of value reuse and get instruction-level parallelism.
Fig.~\ref{fig:mult_clvs_modcode} is the the optimized code, where 16 streams of chained FMA (Fused Multiply-Add) 
operations are executed in a round robin fashion.
Since the FMA latency is 9 and the number of the FMA pipelines is 2, 18 independent operations are required to fill the pipelines.
Therefore, the theoretical efficiency of this code is 89 \% (=16/18).
In addition, the schedule allows for the reuse of register values at short intervals; the number of registers required does not 
exceed 32 (the number of registers in the architecture).
Since each loaded value is reused once, the number of FMAs and loads are equal.
Since the FMA and load pipelines are also equal in number, the load pipeline will not become a bottleneck.
We prevented undesired compiler optimization, which is common subexpression elimination for long-distance reuse, by splitting blocks with if statements.
We also specified the compiler flags that suppress instruction scheduling (\texttt{-Knosch\_post\_ra -Knosch\_pre\_ra -Knoeval}).
We confirmed that these optimizations reduced the number of spills from the original 512 to the optimized 14.

\subsection{Minimizing communication}\label{sec:comm}

The boundary matrices $A_{EO}$ and $A_{OE}$ contain nearest neighbor stencil
communication between surface sites.
To minimize communication time, we adopt the double buffering algorithm 
and implement it by using the uTofu API library to directly use the RDMA engine called 
Tofu Network Interface (TNI) instead of the MPI communication library.
The uTofu interface has lower latency than that of MPI and they can be used together in an application,
so that global operations such as reduction for vector inner-product are processed by MPI functions while the nearest neighbor 1-to-1 communications are done by uTofu.
Choice of a proper rankmap\footnote{We may call it process placement more generally.} and assignments of TNIs are also important to
extract the best performance of TofuD in the neighboring communication.

\begin{figure}[H]
\centering
\includegraphics[width=0.7\linewidth, bb=0 0 517 527]{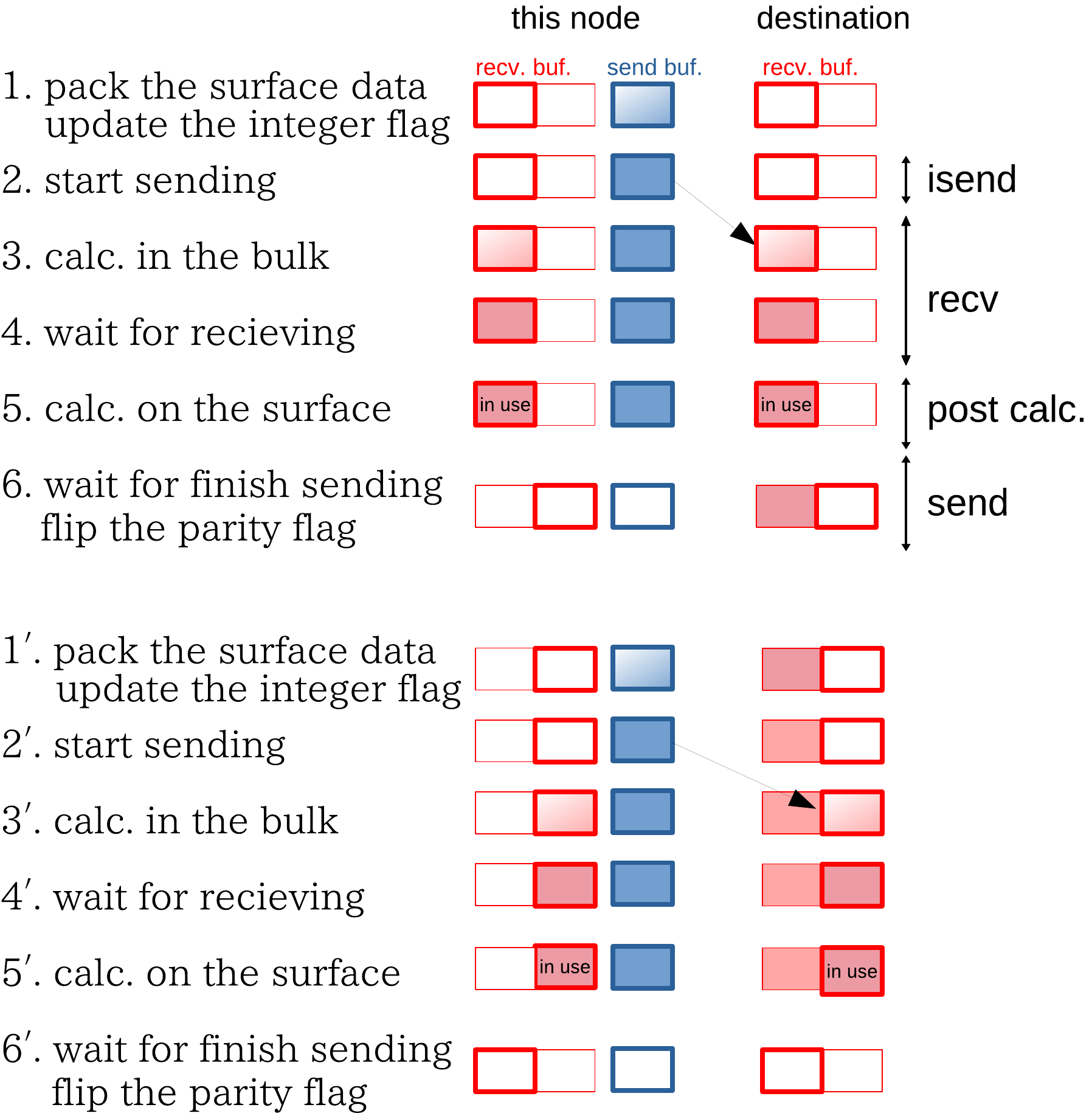} 
\caption{Sketch of double buffering algorithm. The receive buffer is doubled and there is no step to check whether the receive buffer is available. 
 The primed steps use the second receive buffer.
 1--2 and 1${}^\prime$--2${}^\prime$ are for [Start sending...] and 4 and 4${}^\prime$ is for [Wait for receiving...] in Algorithms \ref{alg:qws_prec_s} and \ref{alg:qws_prec_ddd_s}.  Note that the receiver does not clear the integer flag at the end of the buffer but the sender update it before starts sending.
``isend'', ``recv'', and ``send'' correspond to those in Table~\ref{tab:break}.}
\label{fig:double_buffering}
\end{figure}

The double buffering algorithm uses one send buffer and two receive buffers for
each neighboring direction.  The outline of the algorithm is depicted in
Fig.~\ref{fig:double_buffering}.
The RDMA put is issued by \texttt{utofu\_put} instruction.
We use a strong ordering mode with which the arrival of the last byte in the buffer guarantees the arrival of the whole data in the buffer.
The integer valued flag byte at the end of the buffer is updated by the sender.
In order the cache injection mechanism to work, which directly puts the data to the last level cache of the receiver as well as the main memory, 
the receiver should not modify the contents of the receive buffer.
The receiver polls the last byte of the buffer and wait until it has been updated.
After using the arrived data, the parity flag to specify the receive buffer 
out of two buffers is flipped in both sender and receiver sides.
Note that the sender also needs to know to which buffer the data to be sent.
To detect that the sending has been finished, the sender polls 
a queue called Transmit Complete Queue (TCQ) in the RDMA engine.
Another queue to be polled is Message Receive Queue (MRQ), which stores notices related to data receiving.
Although we do not use the information in the MRQ, 
to avoid flooding of the queue we insert
a polling to clean it before waiting the data arrival; since the 
communication takes longer than the calculation in the bulk,
this is the best place to poll the MRQ.

\begin{figure}[H]
\centering
\raisebox{2em}{\includegraphics[scale=0.6, bb=0 0 302 47]{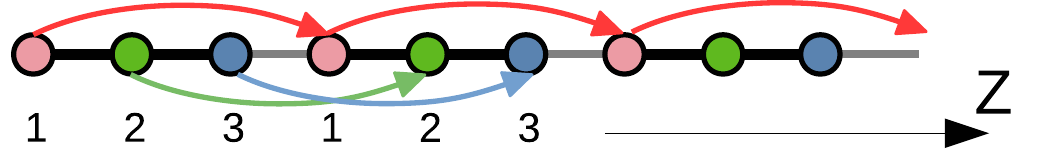}}
\hfil
\includegraphics[scale=0.6, bb=0 0 167 149]{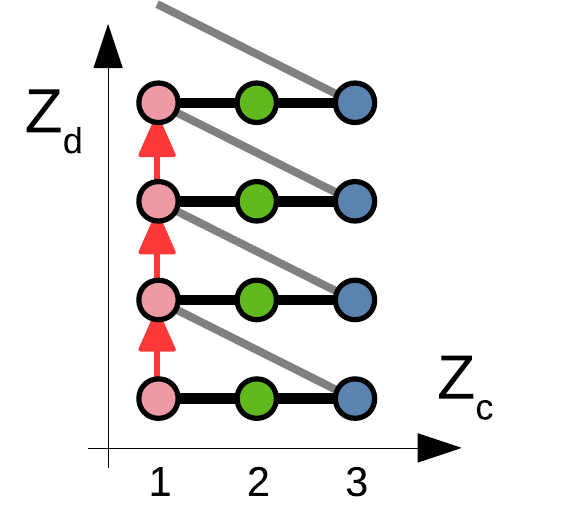} 
\caption{The Z axis (left panel) is divided into a two dimensional mesh-torus topology (the right panel).  Since the original Z axis has a torus topology, 
$\mathrm{Z_d}$ axis is also a torus.}
\label{fig:Zc_and_Zd}
\end{figure}

A natural network topology for LQCD is 4D torus while 
the network of Fugaku is 6-dimensional mesh-torus.
We need a proper rankmap to build a 4-dimensional logical torus,
that is, 4 closed loops of MPI ranks  or MPI processes,
to reduce the hopping in the neighboring communication.
The network of Fugaku has 3 closed loops along X-, Z-, and B-axes of TofuD.  Y-axis is open, A- and C- axes are made of 2 nodes.
We divide Z-axis into two-dimensional mesh-torus, as depicted in Fig.~\ref{fig:Zc_and_Zd}, where $\mathrm{Z_c}$ is continuous open 3 nodes and $\mathrm{Z_d}$ makes a closed loop made of 8 nodes.
Noting that provided 2 independent axes it is easy to construct 
a closed loop topology,
we construct the following 4 logical QCD loops of nodes:
\begin{itemize}
 \item QCD X: 6 nodes, A$\times$ $\mathrm{Z_c}$
 \item QCD Y: 16 nodes, C$\times$ $\mathrm{Z_d}$ 
 \item QCD Z: 24 nodes, X 
 \item QCD T: 64 nodes, B$\times$Y
\end{itemize}
Each node has 4 processes, which are divided into $1\times 2\times 2 \times 1$
along the logical 4D torus.
In total, we have 4D torus of $6\times 32 \times 48 \times 64$ processes
for problem size of $192^4$.

The uTofu interface allows us to specify which TNI to be used
in each communication.  We assign the TNIs such that the communication load is balanced.
In counting the amount of data that the TNI injects, the data to $\mathrm{Z_d}$
axis is weighted by 3 because of its stride in the hopping.
The intra-node commutation is also counted as weight 1,
as we send the data thorough TNI without using the shared memory of the node.

\section{Benchmark results}\label{sec:result}

Benchmark tests are performed on the boost mode 2.2 GHz in which both two floating-point arithmetic pipelines are 
fully utilized at the high frequency\footnote{A64FX has an Eco mode that one of the two floating-point arithmetic pipelines is stopped}.

The elapse time and performance of 500 BiCGStab iterations and 
$A_{EE}$ and $A_{OO}$ regions are listed in Table~\ref{tab:all} and Table~\ref{tab:in} for several problem sizes per MPI process.
We see that the performance of BiCGStab iteration for $32\times 6\times 4\times 6$ is the best in the tests
when the communication is not taken into consideration.

\begin{table}[htbp]
   \centering
   \caption{Elapse time and performance of 500 BiCGSTab iterations measured on two MPI processes using two CMGs
   for several problem sizes per MPI process. FLOP indicates a floating-point operation count calculated theoretically.
   Efficiency indicates floating-point operation efficiency against the single precision floating-point operation peak.
   }
   \label{tab:all}
   \begin{tabular}{ r | r | r | r | r } \hline
Size &  Elapse [s]  & TFLOPS & Efficiency& FLOP \\\hline   
$32\times 6\times 4\times 3$		&0.334867& 	0.8272& 	12.24\%&	69254421000 	\\
$32\times 6\times 4\times 6$		&0.515010& 	1.0839& 	16.04\%&	139560981000 	\\
$32\times 6\times 8\times 6$		&1.145754& 	0.9786& 	14.48\%&	280304661000 	\\
$32\times 6\times 8\times 12$		&2.606202& 	0.8616& 	12.75\%&	561369621000 	\\
$32\times 12\times 8\times 12$	&5.778703& 	0.7773& 	11.50\%&	1122981141000 	\\
\end{tabular}
\end{table}

\begin{table}[htbp]
   \centering
   \caption{Same as Table~\ref{tab:all}, but for regions of $A_{EE}$ and $A_{OO}$ during 500 BiCGSTab iterations.
   }
   \label{tab:in}
   \begin{tabular}{ r | r | r | r | r } \hline
Size &  Elapse [s]  & TFLOPS & Efficiency& FLOP \\\hline   
$32\times 6\times 4\times 3$		&0.068043& 	1.1208& 	16.58\%&	19065600000 \\
$32\times 6\times 4\times 6$		&0.119455& 	1.3170& 	19.49\%&	39329280000 \\
$32\times 6\times 8\times 6$		&0.219403& 	1.4693& 	21.74\%&	80593920000 \\
$32\times 6\times 8\times 12$		&0.559297& 	1.1699& 	17.31\%&	163584000000 \\
$32\times 12\times 8\times 12$	&1.192644& 	1.1146& 	16.49\%&	332328960000 \\
\end{tabular}
\end{table}

Global Allreduce can be a bottleneck when the number of processes is large.
The Allreduce of the MPI implementation for Fugaku is accelerated by using Tofu barrier
which is a offload engine to execute collective operations without involving CPU.
Up to three elements of MPI\_DOUBLE or MPI\_FLOAT, the reduction arithmetics are performed on the Tofu barrier.
It is one of the achievements of co-design that this feature, which was applicable to only one element on K, 
has been extended to three elements on Fugaku, and we stressed the importance of this to the hardware development 
department early in the project.
Taking advantage of this feature, QWS has constructed a solver algorithm as introduced in section \ref{sec:solver}.
To reduce the number and latency of Allreduce for inner product calculation, 
three independent real numbers can be summed up in a single Allreduce operation.

Here we show the benchmark results of the standalone Allreduce function.
In Table~\ref{tab:reduce}, we show Allreduce benchmark results on 72 racks, 27648 nodes of $48\times 12 \times 48$ node shape 
by using ``Intel(R) MPI Benchmarks 2019 Update 6, MPI 1 part (IMB--MPI1)''.
We see that Allreduce up to three elements with the Tofu barrier is about six times faster than one without the Tofu barrier
and it is faster to split MPI\_Allreduce for 15 elements into five MPI\_Allreduce for three elements.

\begin{table}[htbp]
   \centering
   \caption{Allreduce benchmark by IMB--MPI1 on $48\times 12 \times 48$ nodes with and without Tofu barrier.
   Minimum (min), maximum (max), and average (avg) time for repetition number, 10000 are shown.
   The number of bytes (byte) is a messege length to be reduced per one MPI\_Allreuce call.
   And the number of counts (count) is a number of elements. The data type of MPI\_FLOAT is reduced as default of IMB--MPI1.
   }
   \label{tab:reduce}
   \begin{tabular}{ r | r | r | r | r | r | r | r } \hline
    \multicolumn{2}{c}{}  & \multicolumn{3}{|c}{with Tofu barrier}  & \multicolumn{3}{|c}{without Tofu barrier}   \\\hline
    byte & count & min [$\mu$s] & max [$\mu$s] & avg [$\mu$s]& min [$\mu$s] & max [$\mu$s] & avg [$\mu$s] \\\hline
    0&0&		0.09 	&	0.14	&	0.10 &	0.10 &	0.16 &	0.12 \\
    4&1&		7.60 	&	11.33	& 	9.46 &55.69 	&69.05 &	62.83 \\
    8&2&		8.25 	&	10.79&	9.50 &55.79 &68.93 	&62.91 \\
    12&3	&		8.25 	&	10.93&	9.57 &55.89 &69.02 	&62.94 \\
    16&4	&		58.99 &	66.95&	62.68 &56.42 &69.71 	&63.51 \\
    32&8	&		61.50 &	72.34&	66.32 &78.24 &97.57 	&88.14 \\
    64&16&		61.61 &	72.38&	66.31 &78.63 &97.84 	&88.42 \\
    128&	32&	63.70 &	74.45&	68.43 &80.46	&99.56 	&90.10 \\\hline
\end{tabular}
\end{table}

\begin{figure}[htbp]
  \begin{center}
  \includegraphics[width=\textwidth, bb=0 0 340 255]{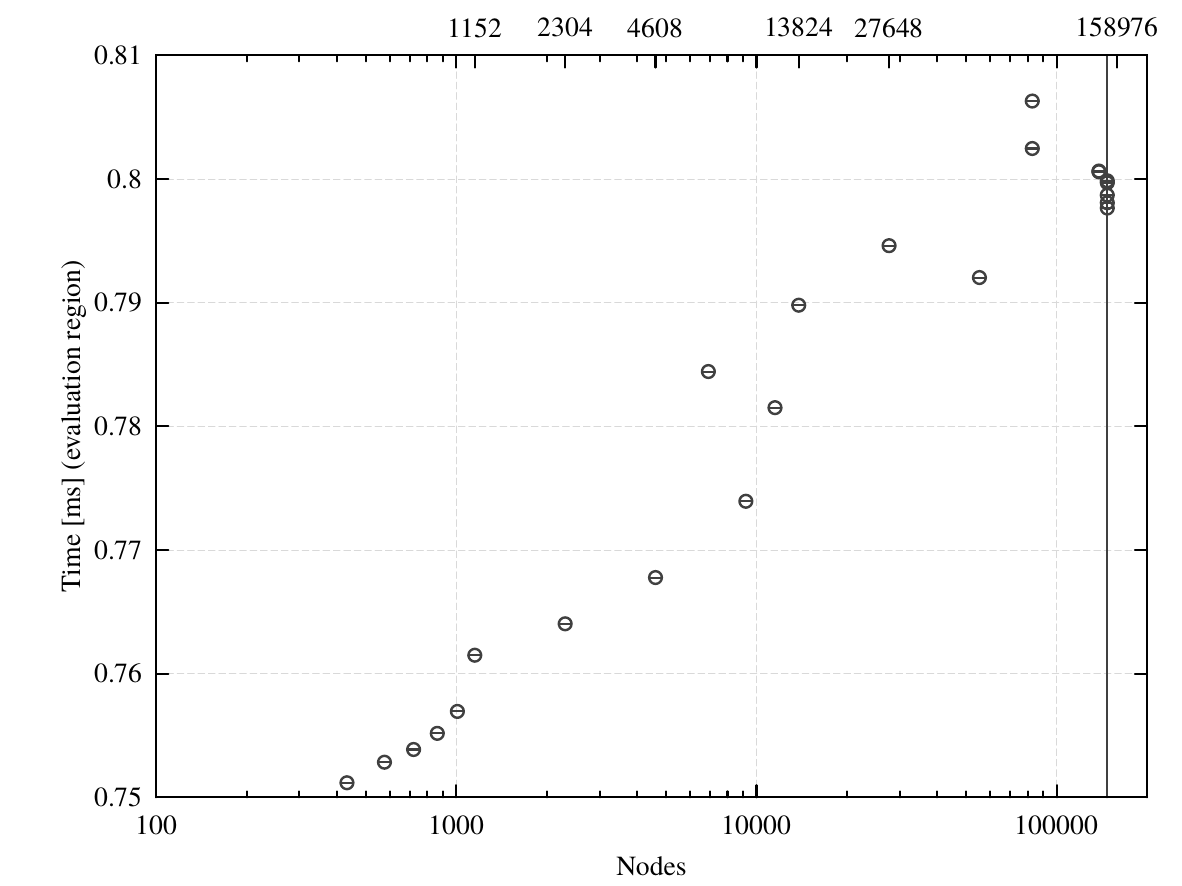}
  \end{center}
  \caption{Weak scaling of the evaluation region. 158976 nodes is a total nodes of Fugaku. 
  The vertical line at 147456 nodes denotes the number of nodes used in the benchmark for the target problem size.}
  \label{fig:weakscaling}
\end{figure}

We show a weak scaling plot of the evaluation region in Fig.~\ref{fig:weakscaling}.
We see a nice weak scaling from 432 nodes to the targeted 147456 nodes with a few exceptions caused by OS jitters.
The elapse time increases $0.05$ ms (about 7 \%) from 432 nodes to 147456 nodes due to the time for Allreduce. 
We can see that the increase time is roughly the same as the Allreduce time in Table \ref{tab:break}.
The elapse times of five benchmark tests on 147456 nodes are
$0.8000$, $0.7998$, $0.7982$, $0.7989$, and $0.7978$ ms, respectively.
These are about 38.3 times faster than the elapse time, $30.65$ ms, for same problem setup on the full system of K.

This performance on Fugaku is 102 PFLOPS sustained with SP arithmetic and corresponds to 10\% efficiency to 
996 PFLOPS of the peak SP performance of 147456 nodes of Fugaku.
Averaged power of the evaluation region is about 20 MW.  The power efficiency is 5 GFLOPS/W.
For comparison, the power efficiency on the LINPACK benchmark at November 2020 is about 15 GFLOPS/W~\cite{top500}.

Table~\ref{tab:break} shows a breakdown of total elapse time for target problem size on 147456 nodes.
The elapse times for ``isend'', ``recv'', and ``send'' are the elapse times for ``isend'', ``recv'', and ``send'' shown in Fig~\ref{fig:double_buffering}, respectively.
The elapse times of ``reduc1'', ``reduc2'', and ``reduc3'' are of 1, 2, and 3 float reductions
in the for loop of iter respectively in Algorithm~\ref{alg:qws_bicgstab_precdd_s}.
Summed total elapse time in the table is slightly longer than $0.8000$ ms which is measured in peak performance tests.
The difference is due to a non-negligible overhead to measure elapse times for each region.
We see the half of time is spent for communication. In the practical production runs, we may better use less nodes.

\begin{table}[htbp]
   \centering
   \caption{Elapse time breakdown. Total time is divided into calculation time (all\_calc) and
   communication time (all\_comm). all\_comm is divided into three parts for
   neighboring communication and three parts for Allreduce.
   }
   \label{tab:break}
   \begin{tabular}{ l | r } \hline
    region & Elapse time [ms]   \\\hline
    all\_calc & 0.400 \\\hline
    all\_comm & 0.407 \\\hline
    isend & 0.029 \\
    recv  & 0.254 \\
    send & 0.062 \\
    reduc1 & 0.015 \\
    reduc2 & 0.016 \\
    reduc3 & 0.031 \\\hline
    total & 0.807 \\\hline    
\end{tabular}
\end{table}

\section{Summary}\label{sec:summary}

We have achieved 102 PFLOPS, 10\% floating-point operation efficiency against single precision floating-point operation peak, 
of Clover--Wilson quark solver on $192^4$ lattice on the supercomputer Fugaku.
Optimization techniques used in QWS are general except for that using ACLE and uTofu
and can be available in other applications.
All the results have been obtained on the evaluation environment in the trial phase.
It does not guarantee the performance, power and other attributes of the supercomputer Fugaku at the start of its public use operation.
However, we expect we shall get better performance with improved compiler, middleware, and McKernel\footnote{No OS jitter is expected.}
which is a light-weight multi-kernel operating system designed for high-end supercomputing~\cite{McKernel}.

\section{Acknowledgment}\label{ackn}

We would like to thank all members of the FLAGSHIP 2020 project, especially 
members of the LQCD working group, and Fujitsu co-design members.
This work is funded by the Japanese Ministry of Education, Culture, Sports, Science 
and Technology (MEXT) program for the Development and Improvement 
for the Next Generation Ultra High-Speed Computer System, 
under its Subsidies for Operating the Specific Advanced Large Research Facilities,
and supported by MEXT as ``Priority Issue on post-K computer'' (Elucidation of the Fundamental Laws and Evolution of the Universe),
``Program for Promoting Researches on the Supercomputer Fugaku'' (Simulation for basic science: from fundamental laws of particles to creation of nuclei),
and JICFuS.

\appendix
\section{Clover--Wilson Dirac operator}\label{sec:dirac}

Clover--Wilson Dirac operator is 
\begin{equation}
\begin{split}
&D = 1 + C -\kappa H \,,\\
&D = 1 + C -\kappa \sum_{\mu=1}^4 ( T^+_\mu + T^-_\mu) \,,
\end{split}
\end{equation}
where 
\begin{equation}
\begin{split}
&C= \frac{i}{2} \kappa \, c_{\rm SW}\, \sigma_{\mu\nu} F_{\mu\nu}(n) \delta_{m,n}\,, \\
&T^\pm_\mu = (1\mp \gamma_\mu) U_{\pm \mu}(n) \delta_{n,m\pm \hat{\mu}}\,.
\end{split}
\end{equation}
$\kappa$ and $c_{\rm SW}$ are input parameters.
$n$ is 4D, $n=(n_1,n_2,n_3,n_4),n_\mu\in\mathbb{Z}$.
$\hat\mu$ is unit vector of $\mu$, for instance, $\hat{1}=(1,0,0,0)$.
$U_{\mu}(n)$ is SU(3) matrix.
$U_{-\mu}(n)=U^\dagger_{\mu}(n-\hat{\mu})$.
$\sigma_{\mu\nu} = i/2 (\gamma_\mu \gamma_\nu - \gamma_\nu \gamma_\mu)$.
$3\times 3$ matrix valued $F_{\mu\nu}(n)$ is a field strength.
$C$ is a block diagonal matrix made by $12\times 12$ Hermitian matrices.
$C_\mathrm{inv}$ appeared in section \ref{sec:solver} is defined as
\begin{equation}
C_\mathrm{inv}= (1 + C)^{-1}\,.
\end{equation}
Multiplication of $H$ and vector is an eight points stencil calculation
for multiplication between $3\times 3$ matrix and four vectors of three elements, i.e. 4 spinor and 3 color.
Gamma matrices $\gamma_\mu$ are
\begin{eqnarray}
\gamma_1
& = &
\left(
\begin{array}{rrrr}
0 & 0 & 0 &+i \\
0 & 0 &+i & 0 \\
0 &-i & 0 & 0 \\
-i& 0 & 0 & 0 \\
\end{array}
\right)
\\[1mm]
\gamma_2
& = &
\left(
\begin{array}{rrrr}
0 & 0 & 0 &+1 \\
0 & 0 &-1 & 0 \\
0 &-1 & 0 & 0 \\
+1& 0 & 0 & 0 \\
\end{array}
\right)
\\[1mm]
\gamma_3
& = &
\left(
\begin{array}{rrrr}
0 & 0 &+i & 0 \\
0 & 0 & 0 &-i \\
-i& 0 & 0 & 0 \\
0 &+i & 0 & 0 \\
\end{array}
\right)
\\[1mm]
\gamma_4
& = &
\left(
\begin{array}{rrrr}
+1& 0 & 0 & 0 \\
0 &+1 & 0 & 0 \\
0 & 0 &-1 & 0 \\
0 & 0 & 0 &-1 \\
\end{array}
\right)
\end{eqnarray}
This is one of the unitary equivalent representations of Gamma matrices.




\bibliographystyle{elsarticle-num}
\bibliography{<your-bib-database>}

\begin{thebibliography}{00}


 \bibitem{Richard2013CoDesign}
   R.~F.~Barrett, {\it et al.}
   On the Role of Co-design in. High Performance Computing. In Advances in Parallel Computing, Volume 24: Transition of HPC Towards Exascale Computing. 141 – 155, 2013.

 \bibitem{Sato2020CoDesign}
  M.~Sato {\it et al.}, Co-Design for A64FX Manycore Processor and ``Fugaku'', SC20, 2020.

 \bibitem{top500}
  TOP500, https://www.top500.org

 \bibitem{A64FX}
 A64FX, Fujitsu's Arm microprocessor which conforms to Armv8 + SVE
 (https://github.com/fujitsu/A64FX).
 
 \bibitem{TofuD}
 Y.~Ajima, {\it et al.}, The Tofu Interconnect D, IEEE Cluster 2018, 2018.

 \bibitem{QWS}
 Y.~Nakamura, Y.~Mukai, K.-I.~Ishikawa, I.~Kanamori,
 Lattice quantum chromodynamics simulation library for Fugaku and computers with wide SIMD
 (https://github.com/RIKEN-LQCD/qws).
 
 \bibitem{Kanamori2021}
 I.~Kanamori {\it et al.}, in preparation (Lattice 2021).
 
 \bibitem{SAP}
 M.~L\"uscher, 
 Lattice QCD and the Schwarz alternating procedure,
 JHEP { 0305}, 052 (2003); Comput. Phys. Commun. { 165}, (2005) 119.

\bibitem{MixedPrecision}
A.~Buttari {\it et al.}, Using mixed precision for sparse matrix computations to enhance the performance while achieving 64-bit accuracy,
ACM Trans. Math.  Soft., 34 (2008) 1,
https://doi.org/10.1145/1377596.1377597.
%
\bibitem{MixedPrecision2}
Y. Saad, 
A Flexible Inner-Outer Preconditioned GMRES Algorithm,
SIAM Journal on Scientific Computing 14, 2, (1993) 461.
\bibitem{MixedPrecision3}
J. A. Vogel,
Flexible BiCG and flexible Bi-CGSTAB for nonsymmetric linear systems,
Applied Mathematics and Computation 188, 1 (227) 226.

\bibitem{flexiblesolver}
 H. Tadano and T. Sakurai,
 On Single Precision Preconditioners for Krylov Subspace Iterative Methods,
 LSSC'07, Lec.  Notes Comput. Sci. 4818 (2008) 721.

 \bibitem{ACLE}
 Arm C Language Extensions,
 https://developer.arm.com/architectures/system-architectures/software-standards/acle

 \bibitem{McKernel} McKernel,
 https://github.com/RIKEN-SysSoft/mckernel


 \end{thebibliography}



\end{document}